\documentclass[reprint,prl,amsmath,amssymb,superscriptaddress]{revtex4-2}
\usepackage{hyperref}
\usepackage{graphics}
\usepackage{graphicx}
\usepackage{enumerate}
\usepackage{hyperref}
\usepackage{braket}
\usepackage{mathtools}
\usepackage{float}
\usepackage{placeins}
\usepackage{xcolor}

\usepackage{dcolumn}
\usepackage{multirow}
\usepackage{amssymb}
\usepackage{perpage}	 

\RequirePackage{lineno} 

\begin{document}

\title{Towards computer-assisted design of hole spin qubits in quantum dot devices}
\author{Ana Ciocoiu}
\affiliation{Department of Electrical and Computer Engineering. University of British Columbia, Vancouver, B.C., Canada.}
\affiliation{Blusson Quantum Matter Institute, University of British Columbia, Vancouver, B.C., Canada.}
\author{Mohammad Khalifa}
\affiliation{Department of Electrical and Computer Engineering. University of British Columbia, Vancouver, B.C., Canada.}
\affiliation{Blusson Quantum Matter Institute, University of British Columbia, Vancouver, B.C., Canada.}
\author{Joe Salfi}
\email{jsalfi@ece.ubc.ca}
\affiliation{Department of Electrical and Computer Engineering. University of British Columbia, Vancouver, B.C., Canada.}
\affiliation{Blusson Quantum Matter Institute, University of British Columbia, Vancouver, B.C., Canada.}
\date{\today}

\begin{abstract}
The design of scalable quantum computers will benefit from predictive models for qubit performance that consider the design and layout of the qubit devices. This approach, has recently been adopted for superconducting qubits, but has received little attention for spin qubits in semiconductors. Here, we employ models for both the device and the quantum mechanical states in the valence band to theoretically investigate the properties of hole spin qubits in laterally gated quantum dots in group-IV materials, which have received significant recent attention. We find that device design impacts qubit properties in unexpected ways. First, the presence of optimal operation points where coherence times are long and qubits can be rapidly manipulated results not only from gate-voltage-induced changes to the Rashba coefficient, but also from gate-voltage-induced changes to quantum dot radius in real devices, which impacts g-factor even when inversion symmetry is preserved (Rashba coupling is zero). Second, the qubit electric manipulation is substantially higher in the realistic anharmonic potential, by an order of magnitude, in the device design we consider, compared with a harmonic potential assumption, because the transverse electric dipole of the qubit depends on electrically mediated couplings to many excited states. Finally, we show that the rapid electric drive, compatible with long coherence times, can be achieved in the single-hole regime, with and without strain. These results establish the need for realistic description of both the device and the quantum mechanical states to support the design of spin qubit devices, identify new ways to control hole qubit properties.
 \end{abstract}

\maketitle


Classical electronic circuit design employs simplified models extracted from device-level simulations. Similarly, the design of large-scale quantum circuits will benefit from a thorough understanding of how device design impacts quantum circuit performance. This approach is beginning to be adopted within the superconducting qubit community, including electromagnetic based extraction of parameters in quantized circuit models in superconducting devices, and development of design frameworks\cite{Nigg2012,yan2020,Minev2021a,Minev2021b}. The understanding of spin qubits in semiconductors\cite{Loss}, which is at an earlier stage of development\cite{Friesen2003,Mohiyaddin2019,Mohiyaddin2020}, could benefit from a similar approach, given recent rapid experimental progress \cite{Watson2018,Hendrickx2}.  

Spin qubits, implemented as quantum dots (QDs)\cite{Veldhorst1,Voisin2016,Maurand,Veldhorst2015a,Watson2018,Hendrickx2,Hendrickx1,Watzinger,Jirovec,Hendrickx2,Hendrickx3} or dopants\cite{Muhonen2014,Kobayashi, Joost} in group-IV materials, are of interest because they utilize the most common materials in the semiconductor industry, and because isotope purification in group-IV materials permits the elimination of the nuclear spin bath, enabling long spin coherence times\cite{Itoh,Muhonen2014,Veldhorst1}. Spin qubits based on valence band holes have recently attracted significant attention in Si\cite{Maurand,Voisin2016,Kobayashi, Joost} and Ge\cite{Hendrickx1,Watzinger,Jirovec,Hendrickx2,Hendrickx3}. This is due to the strong spin-orbit interaction (SOI) experienced by spins in the valence band, which enables electric qubit manipulation through, \textit{e.g.}, electric dipole spin resonance (EDSR)\cite{Bulaev2007}, as well as long-ranged two-qubit operations mediated by superconducting resonators\cite{Kloeffel2013} or mutual capacitances\cite{Salfi}. Holes also lack the valley-orbit degree of freedom that complicates operation of electron spin qubits in group-IV materials\cite{Veldhorst2,Wang1}. Moreover, the $p$-type character of the orbitals of valence band holes suppresses nuclear-spin-induced decoherence compared to conduction band electrons\cite{Testelin2009}. 

For hole spin qubits, the presence of Rashba SOI that enables the desirable electric control generally ties their properties to electric potentials, and makes them more susceptible than electron spin qubits to decoherence and relaxation from electric fluctuations and phonons. Indeed, reported $T_2^*$ coherence times for hole spin QDs lies within 0.13-0.8 $\mu$s\cite{Hendrickx1,Watzinger,Hendrickx3}, compared to $\approx$ 120 $\mu$s \cite{Veldhorst1} for electrons, while reported Hahn-echo coherence times $T_2$ range from 0.2-1.9 $\mu$s \cite{Hendrickx3,Maurand} for holes, compared to $\approx$ 30~ms for electrons\cite{Veldhorst1}. Theoretical studies have identified sweet spots combining fast electric operation and long coherence and relaxation times for hole spins bound to Si:B dopants and QDs in group-IV materials\cite{Salfi, Wang1, Loss1PRXQuantum, Loss2PhysRevLett}, and hole spin coherence times rivaling electron spin in Si has been observed experimentally in Si:B dopants\cite{Kobayashi}. Understanding achievable coherence times both experimentally and theoretically for hole QDs is immensely important because coherence times set upper bounds on quantum operation fidelities (e.g. for two-qubit gates), and increase the time between costly error correction cycles anticipated in future fault tolerant quantum computers. To date, models that account for realistic properties of holes in the valence band do not take into account how devices are built\cite{Bulaev2007,Salfi, Wang1, Loss1PRXQuantum, Terrazos, Loss2PhysRevLett}. Qubit models that incorporate a realistic description of the device and the valence band have yet to be reported. 



In this paper, we theoretically investigate the properties of QD based hole spin qubits in realistic gate-based devices, under different conditions for strain and applied magnetic field direction. We compare the results obtained for potentials in realistic devices with potentials used in previous work. We find that dephasing from electric fields and electric qubit drive in real devices is governed by two additional aspects of spin-orbit coupling relevant to electronic devices that have yet to be identified in the literature. First, the control of the Larmor frequency by varying a gate voltage occurs not only because of changes in the Rashba spin-orbit coefficient induced by the gate\cite{Marcellina2017,Wang1}, but also because of changes to the QD radius induced by the gate, which modifies the g-factor even in zero electric field, when there is inversion symmetry and no Rashba spin orbit interaction. We nevertheless find gate voltages where the Larmor frequency is stable to electric fluctuations, and where the qubit can be driven electrically due to a large transverse electric dipole\cite{Salfi, Wang1, Loss1PRXQuantum, Loss2PhysRevLett}. These optimal operation points are found for both out-of-plane and in-plane applied magnetic fields, and between unstrained and highly strained (0.6\% compressive strain) substrates. Second, the rate of qubit manipulation for EDSR is very sensitive to the anharmonicity of the electric potential. This arises from changes in the excited state spectrum, the states responsible for mediating the electric drive. We find that taking the anharmonicity into account enhances the EDSR rate by 40x compared to a harmonic potential with the same curvature for the device geometries we consider. We also find the qubit manipulation rate to be higher for the in-plane magnetic field than the out-of-plane magnetic field, for fixed Larmor frequency. Finally, we consider the charging of the quantum dot, showing that that optimal operation point can be realized in the single hole regime by an appropriate choice of gate voltages. Our study reveals that the out-of-plane electric field and the occupation of holes are nearly independently controlled by voltage differences and voltage averages, respectively, of plunger and barrier gates in the gated QD. 

\begin{figure}
    \centering
    \includegraphics [width=1\columnwidth]{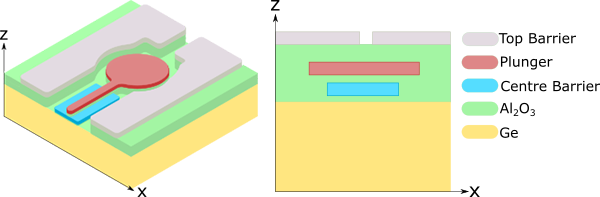}
    \caption{Isometric (left) and front view (right) of the single QD platform build using a Ge substrate. Gates are contained within a 30~nm thick Al$_2$O$_3$ dielectric, which has been partially made transparent for clarity. Spacing is 2.5~nm between Al$_2$O$_3$ interface and centre barrier, 2.5~nm between centre barrier and plunger, and 5~nm between plunger and top barrier.}
    \label{fig:Fig1}
\end{figure}

Although our model works for Si and Ge, we consider for specificity Ge QDs defined by the gate pattern shown in Fig. \ref{fig:Fig1}. The structure consists of three gate layers separated by Al$_2$O$_3$ dielectric. The plunger gate diameter is 90~nm. The Al$_2$O$_3$ is 30~nm thick, with 15 nm between the gate and the active region of the device, 2.5~nm of the dielectric spacing between the plunger and the centre barrier. A hole QD forms beneath the plunger gate within the Ge when a negative voltage is applied to the plunger. Positive voltages applied to the barrier gates create a barrier to a reservoir (not shown). We assume the substrate growth direction is along $\hat{z} \parallel (001)$. \\

\begin{figure*}
\includegraphics[width=2\columnwidth]{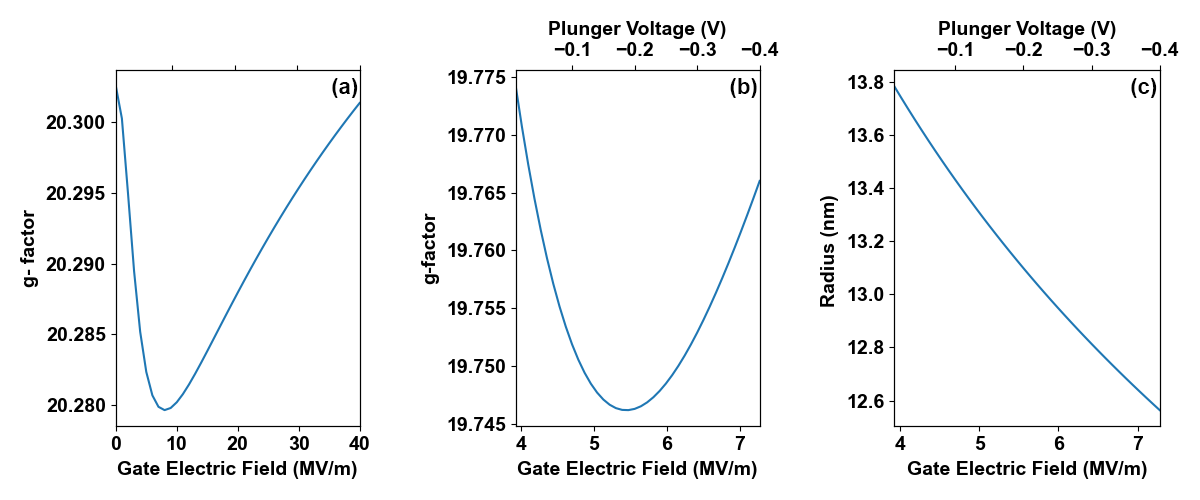}
\caption{\label{fig:wide} Comparison of extracted g-factor for a QD system with toy potential and fixed radius of 16~nm (a) against realistic gate potential with barrier voltage of 1.06V (b). QD radius variation as a function of applied voltage (same barrier voltage as (b) ) and average z-directed field, $Ez$ is shown in (c). All simulations use a zero-strain Ge substrate 15.4 nm thick and out-of-plane magnetic field of 22 mT. We can notice that the 'sweet spot' lies in approximately the same range of electric field values in both (a) and (b), despite significant radius variation across that range for the realistic potential case. }
\label{fig:Fig2}
\end{figure*}

We obtain the eigenstates of the QD using the three-dimensional Kohn-Luttinger Hamiltonian for holes in the valence band. The valence band has $p$-like symmetry ($L=1$). Heavy hole (HH) states are the states with $J=3/2$ and projected angular momentum $m_J=\pm3/2$ and light hole (LH) states are the states with $J=3/2$ and projected angular momentum $m_J=\pm1/2$. Split-off hole states ($J = 1/2$) are not considered in this work, which can be justified in Ge because the split-off valence band is 300~meV away. 
The Hamiltonian describing a single hole QD is:
\begin{equation} \label{eq:1}
H=H_{LK}+H_\epsilon+H_Z+U(\mathbf{r})
\end{equation}
Here, $H_{LK}$ is the Luttinger-Kohn Hamiltonian for the valence band\cite{Luttinger1955}, which accurately decribes both QDs\cite{Bulaev2007} and acceptor dopants\cite{Bir,Bir2},
\begin{equation}
H_{LK}= \begin{bmatrix} P+Q & 0 & L & M \\ 0 & P+Q & M^* & -L^* \\ L^* & M & P-Q & 0 \\ M^* & -L & 0 & P-Q \end{bmatrix},
\end{equation}
where $P=\frac{\hbar^2\gamma_1}{2m_0}(k^2+k_z^2)$, $Q=\frac{-\hbar^2\gamma_2}{2m_0}(2k_z^2-k^2)$, $L=\frac{- \sqrt{3}\hbar^2\gamma_3}{2m_0}k_{-}k_z$, $M=\frac{- \sqrt{3}\hbar^2}{2m_0}((\frac{\gamma_2+\gamma_3}{2})k_{-}^2+(\frac{\gamma_3-\gamma_2}{2})) k_{+}^2$, and $m_0$ is the free electron mass, $\gamma_1$, $\gamma_2$, $\gamma_3$
are Luttinger band-structure parameters with values of 13.15, 4.4, and 5.69 respectively\cite{PhysRevB.4.3460}. The in-plane wave vector $k$ is given by $k^2=k_x^2+k_y^2$ and $k_{\pm} = k_x \pm i k_y$. $H_{LK}$ is written in the basis of HH and LH states:
\begin{equation}
\left\{%
\ket{\tfrac{3}{2},\tfrac{3}{2}}\ket{\tfrac{3}{2},-\tfrac{3}{2}}\ket{\tfrac{3}{2},\tfrac{1}{2}}\ket{\tfrac{3}{2},-\tfrac{1}{2}}
\right\}%
\end{equation}
$H_\epsilon$ is a Bir-Pikus Hamiltonian for strain\cite{Bir,Bir2}, $U(\mathbf{r})$ is the electric potential energy, and $H_Z$ is the Zeeman coupling between the hole and external magnetic field $\mathbf{B}$, given by $H_Z=\mu_B \left(g_1(J_xB_x + J_yB_y + J_zB_z)+g_3(J_x^3B_x+J_y^3B_y+J_z^3B_z)\right)$, $g_1, g_3$ are the linear and cubic Lande g-factors, $\mu_B$ is the Bohr magneton. 
 
The voltages chosen for the gates define the potential energy $U(\mathbf{r})$ for the QD, which is obtained from a finite element electrostatic calculation employing the gate pattern (see Fig. \ref{fig:Fig1}). The relative dielectric permittivity of different materials is taken into account.
We solve the full Hamiltonian numerically in the finite difference approximation, using the external potential $U(\mathbf{r})$, to find the eigenstates $\psi_i$, which are spin-3/2 spinors, and their energies.
The numerical solver uses infinite well boundary conditions. Calculations are presented for unstrained and strained Ge layers of 15.4 nm depth. 
The two lowest-energy eigenstates are, as expected, mainly $+3/2$ and $-3/2$ heavy-hole states.

The coupling of spin qubits to electric fields is described by the longitudinal and transverse spin-electric dipoles, $v$ and $\chi$, respectively. These electric dipoles are constructed from integrals $p_{ijk}$,
\begin{equation}
p_{ijk}=\int \psi_i(r)^*qx_k\psi_j(r) d^3r.
\end{equation}
that express couplings between states $i$ and $j$ due to an electric field along the $k$ axis. The transverse electric dipole $v=p_{12k}$ determines the Rabi frequency $\tau_{\rm Rabi}^{-1}=vE_{ac}/h$ of the oscillation between qubit states $i=1$ and $j=2$ due to an applied electric field of magnitude $E_{ac}$, oscillating at the qubit frequency, along the $k$ axis in space. The electric dipole matrix elements with $i = j$ describe the shift in energy of state $i$ due to an electric field $E$, and the longitudinal electric dipole $\chi=p_{22k}-p_{11k}$ determines the change in qubit splitting $\Delta \varepsilon=\chi E$ due to electric noise, which causes pure dephasing in the presence of charge noise (fluctuating electric fields).

The eigenstates $\psi_i$ are computed numerically under two different assumptions: first, using harmonic potentials that ignore gate voltage induced changes in QD radius, and second, using potentials predicted from voltages applied to the gates in Fig. \ref{fig:Fig1}. For the first, a 16~nm QD is modeled with a laterally harmonic potential and independently tunable vertical electric field $E_z$, $U(\mathbf{r})=\frac{1}{2}m_p {\omega_0}^2(x^2+y^2)+qE_zz$ , where $m_p= m_0/(\gamma_1 + \gamma_2)$ is the in-plane effective mass of the heavy holes and $\omega_0=\hbar/(m_p a_0^2)$ is the harmonic oscillator frequency, and $a_0$ is the quantum dot radius. The Hamiltonian is solved numerically and the resulting g-factor, defined by the energy difference between the two qubit states $g\mu_B B$, is plotted as a function of vertical field $E_z$ in Fig. \ref{fig:Fig2}(a). The result of our finite numerical solution to the Kohn-Luttinger equations is in qualitative agreement with the variational calculation in ref.~\cite{Wang1}. That is, we observe a local minimum in the g-factor, occurring here at $E_z$=8~MV/m associated with the vanishing of the longitudinal electric dipole that causes pure dephasing from electric noise. This non-monotonic behavior of $g$ as a function of the gate field was previously identified to arise from the non-monotonic behaviour of the Rashba SOI coefficient\cite{Marcellina2017, Wang1}. 

The effective qubit g-factor and QD radius for the real device potential are plotted as a function of plunger gate voltage and average electric field in Fig.~\ref{fig:Fig2} (b),(c). We define the average electric field as $\langle E_z \rangle = \langle \Psi | -(d/dz) U(\mathbf{r}) | \Psi \rangle$ since the hole state will not experience a perfectly uniform electric field $E_z$ in realistic devices. Similar to Fig \ref{fig:Fig2}(a), we observe a local minimum in the g-factor, this time at an electric field 5.45~MV/m. In both cases, as expected, the qubit basis states are found to be composed of mainly HH states (albeit with non-zero LH-HH mixing), such that the system is operating near the 2-D limit. 

The observation of the local minimum at 5.45 MV/m instead of 8 MV/m means that the optimal operation point has shifted upon consideration of the realistic device potential. We have investigated the origin of the shift, and found it to be as follows. In the perfect 2D limit, the g-factor changes with electric field because of the dependence of the Rashba SOI on  electric field. In the case of a QD with in-plane confinement, changes in the in-plane QD radius change the QD's effective g-factor at fixed electric field, even when there is inversion symmetry (in zero electric field). We observe in the device potential that increasing gate voltage modifies not only the Rashba electric field, as previous identified, but also reduces the QD radius (Fig.~\ref{fig:Fig2}(b),(c)). In comparison, the QD radius is assumed to be fixed in the simplified potential (Fig~\ref{fig:Fig2} (a)). While the gate voltage modifies both the Rashba SOI coefficient and the QD radius, the change in the Rashba SOI has more influence on the overall curve shape, producing the minimum of the g-factor with the gate voltage in Fig. \ref{fig:Fig2}(b) at 5.45~MV/m.

\begin{figure}
    \centering
    \includegraphics [width=0.95\columnwidth]{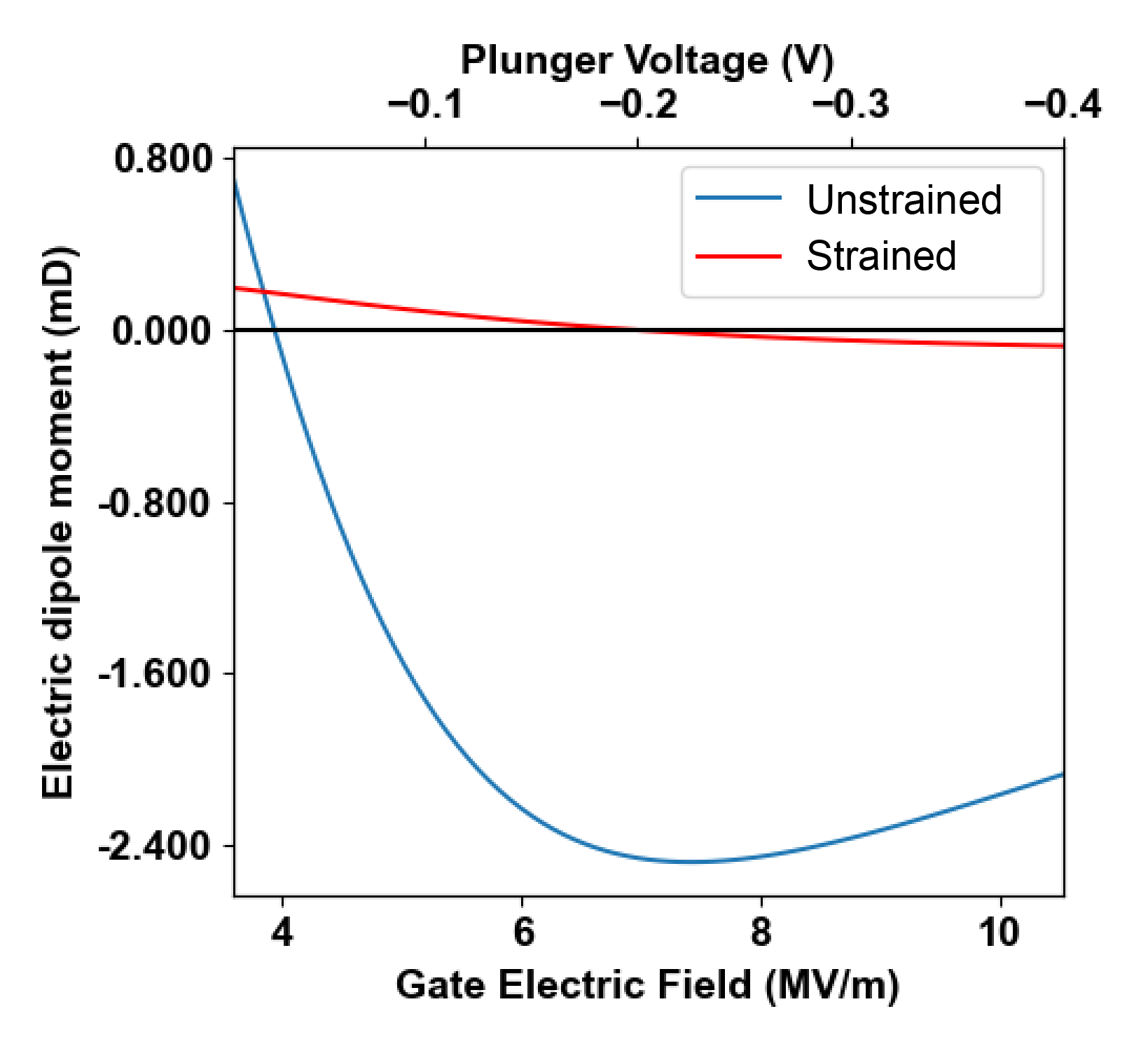}
    \caption{Plot of the longitudinal dipole in Debye for both unstrained ($\epsilon_x, \epsilon_z = 0$) and strained ($\epsilon_x= -0.0063, \epsilon_z  = 0.0044$) Ge. Increased electric field dependence of the dipole in the unstrained case can be seen. Note the existence of an x-axis crossing in both cases.}
    \label{fig:FigStrain}
\end{figure}

We now calculate the longitudinal electric dipole for the qubit states $\psi_i$ in the real device, for two different values of strain, completely unstrained materials, and strain of $-0.6$~\% similar to the value expected for Ge quantum wells in the literature\cite{Hendrickx1}. The results for the longitudinal electric dipole are shown in Fig.~\ref{fig:FigStrain}, which is found to have stronger variation with electric field in the unstrained material. Similarly, the transverse dipole is about 5x larger than in the strained Ge (results not shown). Importantly, there exists a choice of gate voltage (electric field) where the longitudinal dipole vanishes in both cases, demonstrating the existence of an optimal operation point where dephasing vanishes to first order in electric field noise amplitude. In other words, the presence of optimal operation points is not tied to whether or not the material is strained. 

\begin{figure}
    \centering
    \includegraphics [width=0.95\columnwidth]{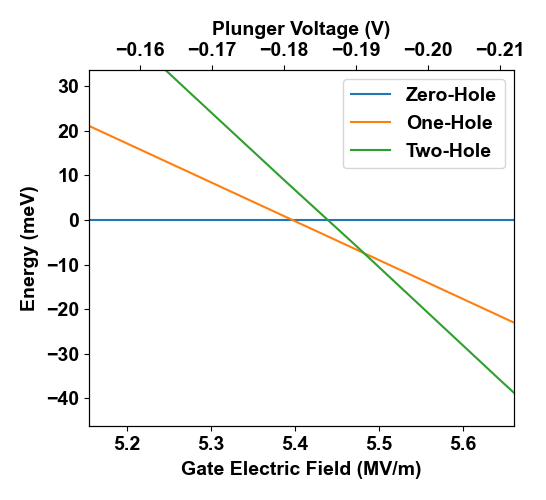}
    \caption{Plot of the zero, one, and two-hole regimes as a function of realistic gate potential and $\langle Ez \rangle$ for a barrier voltage of 1.06V. Note that the one-hole regime resides in the plunger voltage range of $\thicksim$0.181-0.191 V, giving the region a width of $\thicksim$10 mV. The sweet spot of Fig. \ref{fig:Fig2}(b) resides within this one-hole regime. Out-of-plane magnetic field is 22mT, and the substrate is 15.4 nm thick unstrained Ge.}
    \label{fig:Fig3}
\end{figure}

Having found the optimal gate voltages where the longitudinal dipole vanishes, we show that this can be obtained in the regime where a single hole occupies the QD. To investigate this, we calculated the energy of the zero-hole, one-hole and two-hole states by considering the Coulomb interactions on a single QD\cite{Yang}. We set the zero-hole energy $E_0$ as a reference point: then, the one-hole energy $E_1$ is simply the ground state energy eigenvalue, $\mu_1$, obtained from the numerical solver using the realistic gate potential as $U(\mathbf{r})$. 
The two-hole energy $E_2$ is calculated as $2\mu_1+U_C$, where $U_C$ is the Coulomb repulsion integral between two holes residing in the same QD,
\begin{equation}
U_C= \iint {|\psi({\bf r_1})|}^2 \frac{q}{4\pi \epsilon_0 \epsilon_r|{\bf r_1}-{ \bf r_2}|}{|\psi({\bf r_2})|}^2 d^3{\bf r_1} d^3{\bf r_2}
\end{equation} 
and $\psi$ is the ground-state wavefunction of a single hole obtained from the numerical solver. We have calculated the Coulomb repulsion integral by Monte-Carlo integration. The results for $E_0$, $E_1$ and $E_2$ are plotted as a function of applied gate voltage and average electric field in Fig. \ref{fig:Fig3}. 
The one-hole regime occurs when the energy of the single-hole state is lower than the zero and two-hole states, $E_1 < E_0$ and $E_1 < E_2$. 

To line up the one-hole regime, where $E_1 < E_2$ and $E_1 < E_0$, with the local minimum of the g-factor, we take advantage of the fact that the QD experiences a potential produced by multiple in-plane gates. We employ the intuition that the difference between the plunger and the barrier gate voltages mostly controls electric field $\langle Ez \rangle$ experienced by the hole, which should set the location of the sweet spot, and the mean value of the plunger and barrier controls mainly occupation. Varying the barrier gate voltage, we find that this shifts the plunger voltage where the g-factor minimum is found in Fig. \ref{fig:Fig2}(b), and that the average value $\langle Ez \rangle$ of the electric field where the g-factor minimum is found is roughly constant, in spite of the changing plunger gate voltage. This allows us to line up the local minimum at 5.45~MV/m with the one-hole regime in Fig. \ref{fig:Fig3}.

As a consistency check, we examine the lever-arm of the plunger gate, which indicates the change in electrochemical potential of the QD per unit change in the electrostatic energy of the electrode\cite{Haruki}. The gate lever-arm is calculated using the average value for $U_C$ in (5) over a sweep of the plunger gate voltage at fixed barrier gates voltages, divided by the width of the one-hole regime. It is found to be $\approx 0.72$, reflecting the relatively small dielectric thickness in the device design (Fig.~\ref{fig:Fig1}). As expected, the lever-arm is higher than values from the literature where gates are farther from the well. A lever-arm of $\approx 0.2$ was found for Ge QWs capped with 22~nm GeSi and 20~nm Al$_2$O$_3$\cite{Hendrickx1, Hendrickx3}. Our predicted lever-arm is similar to the lever-arm of 0.85 found for nanowire field-effect transistor based QDs with effective oxide thicknesses of $< 2$~nm \cite{Voisin2016}.
\begin{figure}
    \centering
    \includegraphics [width=0.95\columnwidth]{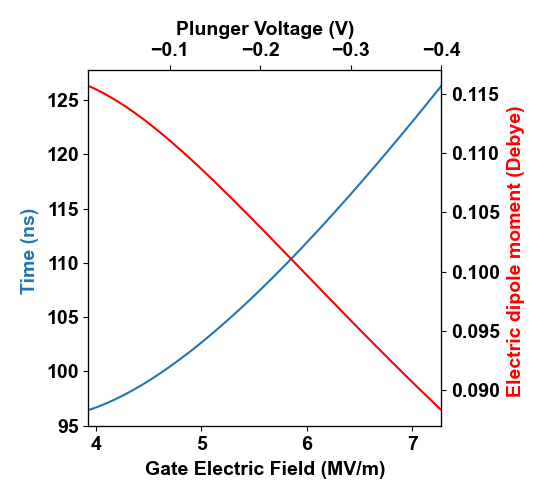}
    \caption{$t_{Rabi}$ (blue) and $v$ (red) plotted as a function of applied gate voltage and average electric field. The value of the driving field $E_{AC}$ is 18 kV/m, and the out-of-plane magnetic field is 22mT. The substrate is 15.4 nm thick unstrained Ge.}
    \label{fig:Fig4}
\end{figure}

We have evaluated the transverse dipole defined in (4) in order to understand the impact of the device on the qubit manipulation rate. The EDSR Rabi time is obtained from the transverse dipole $v$ using
\begin{equation}
t_{Rabi} = \frac{h}{E_{AC}|v|},
\end{equation}
where $E_{AC}$ is the strength of an oscillating in-plane electric field on resonance with the qubit. The trends in $t_{Rabi}$ and $v$ are plotted as a function of applied gate voltage and average electric field in Fig. \ref{fig:Fig4}. 

Comparing these results to the result obtained for the harmonic potential calibrated to the same curvature, we find that both the transverse and longitudinal dipoles are underestimated by the harmonic potential model. Since our harmonic potential is chosen to reproduce the same curvature at the bottom of the anharmonic potential, the difference in the results must be due to the long-distance behaviour of the potential. In particular, for realistic devices, the potential $U(\mathbf{r})$ approaches a finite value far away from the gates, and at this energy, the energy level density becomes very large. This contrasts the harmonic potential typically used to describe the energy spectrum of QDs\cite{Bulaev2007}, where the excited states are spaced equally apart. Though clearly unphysical for sufficiently highly excited states and at large distances where the potential diverges, the harmonic potential is convenient because it is usually believed to describe the lowest energy states well. We find this to not be the case for the longitudinal and transverse dipole. In particular, the change to wavefunctions and energy level densities from the treatment of the anharmonic potential impacts the EDSR matrix elements greatly. In the framework of Schrieffer Wolff transformations \cite{SchriefferW, Winkler}, the higher excited state energies and smaller wavefunctions of the unbounded parabolic potential would be anticipated to reduce the transverse dipole and increase the EDSR time compared to the realistic potential. 

We find the underestimate is a factor of $\thicksim$36 for the transverse dipole, and $\thicksim$3 for the longitudinal dipole, for the device we have considered. This is a remarkable quantitative difference between two approaches that are so similar at first glance, and highlights the importance of considering realistic device potentials to produce proper qubit performance predictions. The increase by 36x of the transverse electric dipole compared to a harmonic potential implies a >1000x increase in long-range two-qubit coupling mediated by electric dipoles, which is relevant to circuit quantum electrodynamics and direct electric dipole-dipole interactions\cite{Kloeffel2013,Salfi}. 

\begin{figure}
    \centering
    \includegraphics [width=1.0\columnwidth]{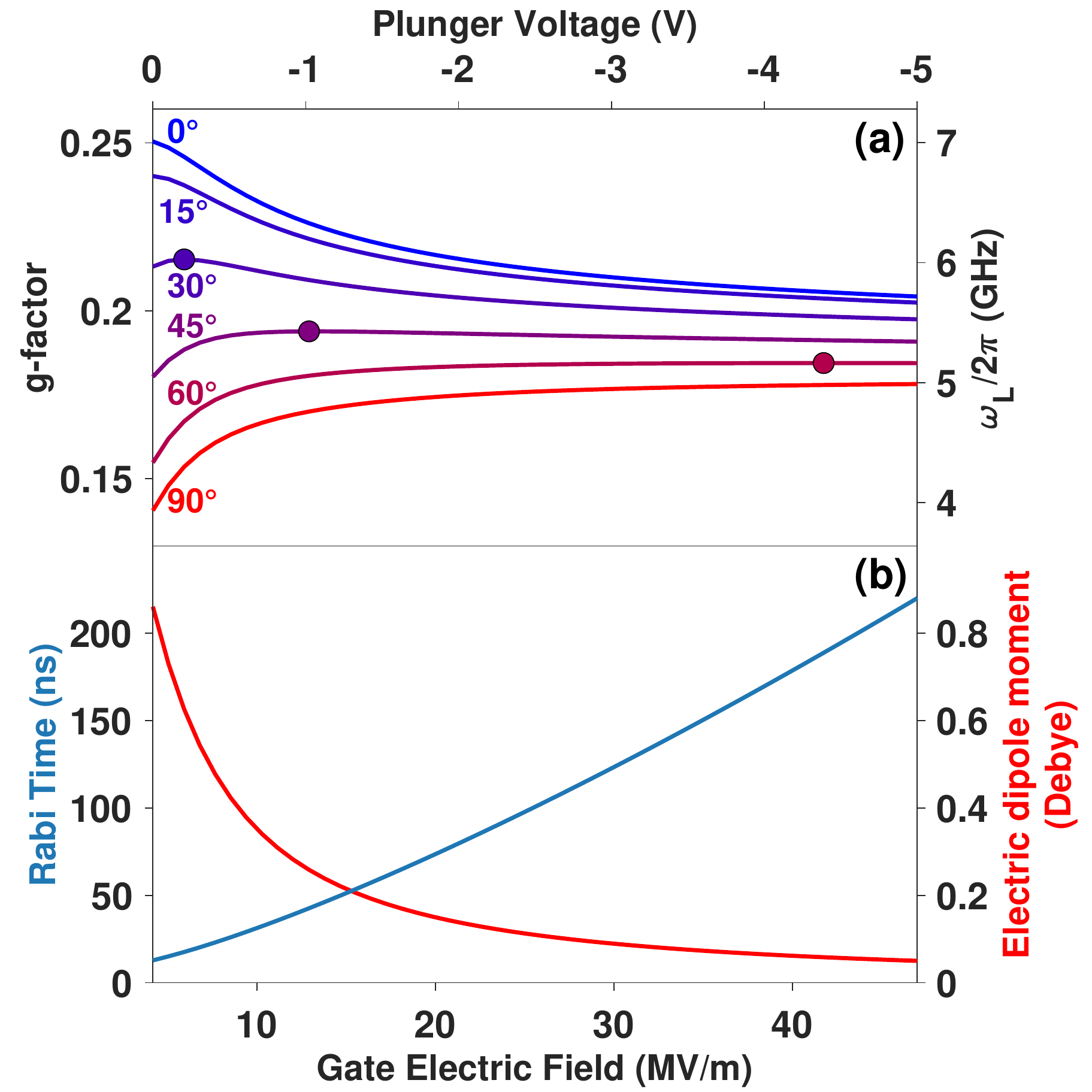}
    \caption{Qubit properties for in-plane magnetic field. (a) Plunger voltage dependence of g-factor and corresponding Larmor frequency, $\omega_L$, for different orientations of magnetic field $B = 2$ T in xy-plane. $B$ is along x-axis (y-axis) at $0^\circ$ ($90^\circ$); blue (red) line. The circles at $30^\circ,45^\circ$ and $60^\circ$ lines highlight the locations of the sweet spot. (b) Plunger voltage dependence of electric dipole moment along x-axis when magnetic field is along y-axis, and corresponding Rabi time for driving field $E_{AC}=18$ kV/m.}
    \label{fig:inPlaneB}
\end{figure}
 
Finally, we investigated the properties of the hole spin qubit for in-plane magnetic fields that can be more readily tolerated by superconducting devices. The in-plane g-factor is around 0.2, as shown in Fig. \ref{fig:inPlaneB}(a), which closely matches the experimentally reported values in planar QDs \cite{Hendrickx1, hendrickx2020fast, Hendrickx2, miller2021effective}. Compared to the out-of-plane g-factor, the small in-plane g-factor requires two orders of magnitude larger magnetic field to operate at the same Larmor frequency. This large anistropy is consistent with previous theoretical predictions \cite{nenashev2003wave, maier2013tunable}, and has been demonstrated experimentally in planar QDs \cite{miller2021effective}. Moreover, the vertical electric field dependence of the in-plane g-factor depends on the magnetic field direction in the xy-plane. In particular, the difference between 0$^\circ$ and 90$^\circ$, two equivalent axes in cubic materials, is due to the absence of perfect circular symmetry in the realistic confining potential. Interestingly, the g-factor sensitivity to the magnetic field direction gives an extra degree of freedom for controlling the plunger voltage and the electric field at which the sweet spot occurs, as illustrated in Fig. \ref{fig:inPlaneB}(a). We emphasize here the importance of the cubic g-factor term for in-plane magnetic field calculations, as the effect of the linear Zeeman term is negligible compared to it. This applies for both the harmonic and the realistic potentials. In contrast, the effective out-of-plane g-factor is totally dominated by the linear Zeeman term. Fig.~\ref{fig:inPlaneB}(b) shows the variation of $v$ and the associated $t_{Rabi}$ versus vertical electric field when $B$ is along y-axis. At low electric field, $v$ goes up to 0.8 Debye, which is about 7 times greater than when $B$ is out of plane for the same Larmor frequency. Therefore, qubits defined by in-plane field can be manipulated faster. 

\textit{Conclusion-} The theoretical model in this paper is the first we are aware of for describing qubits based on hole spin in the valence band that takes into account a realistic description of the device, strain, and a realistic quantum mechanical theory for qubits in the valence band. It is found that (a) optimal operation points can still be identified for realistic devices where gate voltage not only controls the Rashba coefficient, but also changes the QD radius, and that the optimal operation points occur for strained and unstrained devices, (b) the qubit manipulation rate is significantly underestimated by harmonic potentials compared to realistic anharmonic device potentials, (c) in-plane magnetic field direction allows for an extra degree of freedom in tuning the location of the sweet spot and (d) in-plane magnetic field leads to a sevenfold improvement in qubit manipulation time at the same Larmor frequency. The increase by 36x of the transverse electric dipole compared to a harmonic potential implies a $>1000$x increase in long-range two-qubit coupling mediated by electric dipoles, a significant result. From a high level, these results demonstrate a minimal model to evaluate the properties of qubits from device layouts, and underscore the need to describe both the device and quantum mechanical states of the valence band. In the future, this work could be extended to describe multi-qubit circuits.
\\
\begin{acknowledgements} This work was undertaken with support from the Stewart Blusson Quantum Matter Institute, the National Science and Research Council of Canada through the Discovery Grant scheme, the Canadian Foundation for Innovation through the John Edwards Leaders Foundation scheme, and the Canada First Research Excellence Fund, Quantum Materials and Future Technologies Program. AC acknowledges financial support from NSERC Canada Graduate Scholarship and the NSERC CREATE program in Quantum Computing. AC and MK acknowledge financial support from the QMI QuEST fellowship program. The authors acknowledge CMC Microsystems for the provision of computer aided design tools that were essential to obtain the results presented here. This research was supported in part through computational resources provided by Advanced Research Computing at the University of British Columbia. \end{acknowledgements}

%

\end{document}